\documentclass[10pt, final]{article}
\usepackage[top=1in, bottom=1in, left=1in, right=1in]{geometry}
\usepackage{graphicx}
\usepackage{enumerate}
\usepackage{natbib}
\usepackage{url} 
\usepackage{amsmath, amsfonts, scalerel, mathrsfs, amssymb, amsthm,bbm}
\usepackage[titletoc,title]{appendix}
\usepackage{hyperref}
\usepackage{caption}
\usepackage{enumitem}
\theoremstyle{plain}
\newtheorem{theorem}{Theorem}[section]
\newtheorem{prop}[theorem]{Proposition}
\newtheorem{assump[theorem]}{Assumption}
\newtheorem{cor}[theorem]{Corollary}

\theoremstyle{definition}
\newtheorem{definition}[theorem]{Definition}

\DeclareFontFamily{U}{mathx}{\hyphenchar\font45}
\DeclareFontShape{U}{mathx}{m}{n}{
      <5> <6> <7> <8> <9> <10>
      <10.95> <12> <14.4> <17.28> <20.74> <24.88>
      mathx10
      }{}
\DeclareSymbolFont{mathx}{U}{mathx}{m}{n}
\DeclareFontSubstitution{U}{mathx}{m}{n}
\DeclareMathAccent{\widecheck}{0}{mathx}{"71}

\newcommand{\blind}{0}

\def\old{{\em detect-and-forget}}
\def\oldnoitalics{detect-and-forget}
\def\newknownsig{{\em corrected}}
\def\newexact{{\em corrected-exact}}
\def\newest{{\em corrected-est}}
\DeclareMathOperator*{\bigcdot}{\scalerel*{\cdot}{\bigodot}}
\newcommand{\indep}{\mathrel{\text{\scalebox{1.07}{$\perp\mkern-10mu\perp$}}}}
\def\hM{\widehat M}
\def\b0{\mathbf{0}}

\def\be{\mathbf{e}}
\def\bz{\mathbf{z}}
\def\by{\mathbf{y}}
\def\bx{\mathbf{x}}

\def\bu{\mathbf{u}}
\def\bw{\mathbf{w}}
\def\bR{\mathbf{R}}
\def\bnu{\boldsymbol{\nu}}
\def\bmu{\boldsymbol{\mu}}
\def\bbeta{\boldsymbol{\beta}}
\def\bvarepsilon{\boldsymbol{\varepsilon}}

\begin{document}

\def\spacingset#1{\renewcommand{\baselinestretch}%
{#1}\small\normalsize} \spacingset{1}


\if0\blind
{
  \title{\bf Valid Inference Corrected for Outlier Removal}
  \author{Shuxiao Chen \hspace{.2cm}\\
    Department of Statistics, University of Pennsylvania\\
    and \\
    Jacob Bien 
    \\
    Data Sciences and Operations, University of Southern California}
  \maketitle
} \fi

\if1\blind
{
  \bigskip
  \bigskip
  \bigskip
  \begin{center}
    {\LARGE\bf Title}
\end{center}
  \medskip
} \fi

\bigskip
\begin{abstract}
Ordinary least square (OLS) estimation of a linear regression model is
well-known to be highly sensitive to outliers. It is common practice
to (1) identify and remove outliers by looking at the data and (2) to fit
OLS and form confidence intervals and $p$-values on the remaining data
as if this were the original data collected.  This standard ``\oldnoitalics''
approach has been shown to be problematic, and in this paper we highlight
the fact that it can lead to invalid inference and show how recently
developed tools in selective inference can be used to properly account
for outlier detection and removal.
Our inferential procedures apply to a general class of outlier removal
procedures
 that includes several of the most commonly used approaches. 
We conduct simulations to corroborate the theoretical results, and we apply our method to three real data sets to illustrate how our inferential results can differ from the traditional \old~strategy. A companion {\tt R} package, {\tt outference}, implements these new procedures with an interface that matches the functions commonly used for inference with  {\tt lm} in {\tt R}.
\end{abstract}

\noindent%
{\it Keywords:} confidence intervals, linear regression, outlier, $p$-value, selective inference 
\vfill

\newpage
\spacingset{1.45} 


\section{Introduction}
\label{sec:intro}

Linear regression is routinely used in just about every field of science. In introductory statistics courses, students are shown cautionary examples of how even a single outlier can wreak havoc in ordinary least squares (OLS).
Outliers can arise for a variety of reasons, including recording errors and the occurrence of rare phenomena, and they often go unnoticed without careful inspection \citep[see, e.g.,][]{belsley2005regression}.
Given this reality, one simple strategy adopted by practitioners is a two-step procedure which we will refer to as \old:
\setlist{nolistsep}
\begin{enumerate}
  \item detect and then remove outliers;
  \item fit OLS and perform inference on the remaining data \emph{as if this were the original data set}.
\end{enumerate}
While this simple approach is extremely common, there are two major
problems \citep{welsh2002journey}. 
First, accurate detection of outliers can be challenging: 
In the presence of multiple outliers, classical influence measures, such as OLS residuals, Cook's distance \citep{cook1977detection}, and DFFITS \citep{welsch1977linear}, 
can be misleading, leading potentially to missed outliers and falsely detected
outliers \citep[see, e.g.,][for more on ``masking'' and
``swamping'']{hadi1993procedures}. 
This first problem has received considerable attention, leading to the
development of \emph{robust regression} methods, in which one uses
methods that are less sensitive to outliers \citep[see, e.g.,][]{maronna2006robust}. 
 A foundational method in this category is Huber's M-estimator \citep{huber1981robust}, where one minimizes Huber's loss function:
\begin{equation}
\label{eq:huber}
\underset{\beta \in \mathbb{R}^p}{\min} \ \frac{1}{n}\sum_{i=1}^n \rho(y_i - X_{i,\bigcdot} \beta,\lambda) \qquad\text{ where }\qquad \rho(r, \lambda)=
\begin{cases}
  \frac1{2}r^2&\text{ if }|r|\le\lambda\\
  \lambda|r| - \frac1{2}\lambda^2&\text{ if }|r|>\lambda,
\end{cases}
\end{equation}
where $X_{i, \bigcdot}$ is $i$-th row of the design matrix. 
The ``vanilla'' Huber's estimator has been shown to be insufficiently robust 
 \citep{rousseeuw1984least,zaman2001econometric}, 
 but state-of-the-art robust methods do exist, 
such as MM-estimation \citep{yohai1987high}. 

This first problem with \old~has received much attention; the
focus of this paper, however, is on a second problem.
In the second step of the \old~approach, in which one performs downstream
statistical inference based on the refitted OLS estimator,   
we show that the confidence intervals and hypothesis tests have
incorrect operating characteristics.  This second issue is
orthogonal to the first: whether or not one is able to accurately
identify outliers, if one chooses to search for and remove outliers,
one must account for this step when doing subsequent inference.   We
emphasize that our solution to this second problem does not 
address the first problem of accurate outlier detection.  
Given the widespread continued use of classical outlier detection methods, we
develop a practical fix to this second problem, allowing for valid
inference after using the classical outlier detection methods (including OLS residuals,
Cook's distance and DFFITS) or after using Huber's estimator. 

The inferential problem with \old~stems from its use of the same data twice.
While the term ``outlier removal'' might lead one to think of Step 1 as a clear-cut, essentially deterministic step, 
in fact Step 1 should instead be thought of as ``potential outlier removal,'' 
an imperfect process in which one has some probability of removing non-outliers,
a process that can alter the distribution of the data.
The act of searching for and removing potential outliers must be considered as part of the data-fitting procedure and thus must be considered in Step 2 when inference is being performed. 
Similar concerns over ``double dipping'' are well-known in prediction problems,
in which \emph{sample splitting} (into training and testing sets) is a common remedy. 
However, such a strategy does not translate in an obvious way to the outlier problem: suppose one splits the observations into two sets, searching for and removing potential outliers on the first set and then performing inference on the second set of observations.  In such a case, one is of course left vulnerable to outliers in the second set throwing off the inference stage.

The idea of properly accounting for a previous look at the data is known as \emph{selective inference} \citep{yekutieli2012adjusted,taylor2015statistical}. Much recent work is focused specifically on accounting for selection of a set of variables before performing inference \citep{fithian2014optimal,loftus2015selective,lee2016exact,panigrahi2016bayesian}. In our case, the selection is of observations rather than variables, but we show that the machinery of \cite{loftus2015selective,lee2016exact}, namely conditioning on a stochastic selection event, can be naturally adapted to our context.

In fact, illustration of the problem with \old~has appeared in some literature. \citet{rico2018white} showed how the White test for heteroscedasticity can fail using the \old~approach under asymmetric errors;
however, when the errors come from a symmetric distribution, they
show how the theory of \citet{rico2017marked} 
can lead to the \old~approach being valid asymptotically and having good finite-sample performance.

\begin{figure}[t]
\centering
\includegraphics[width = \textwidth]{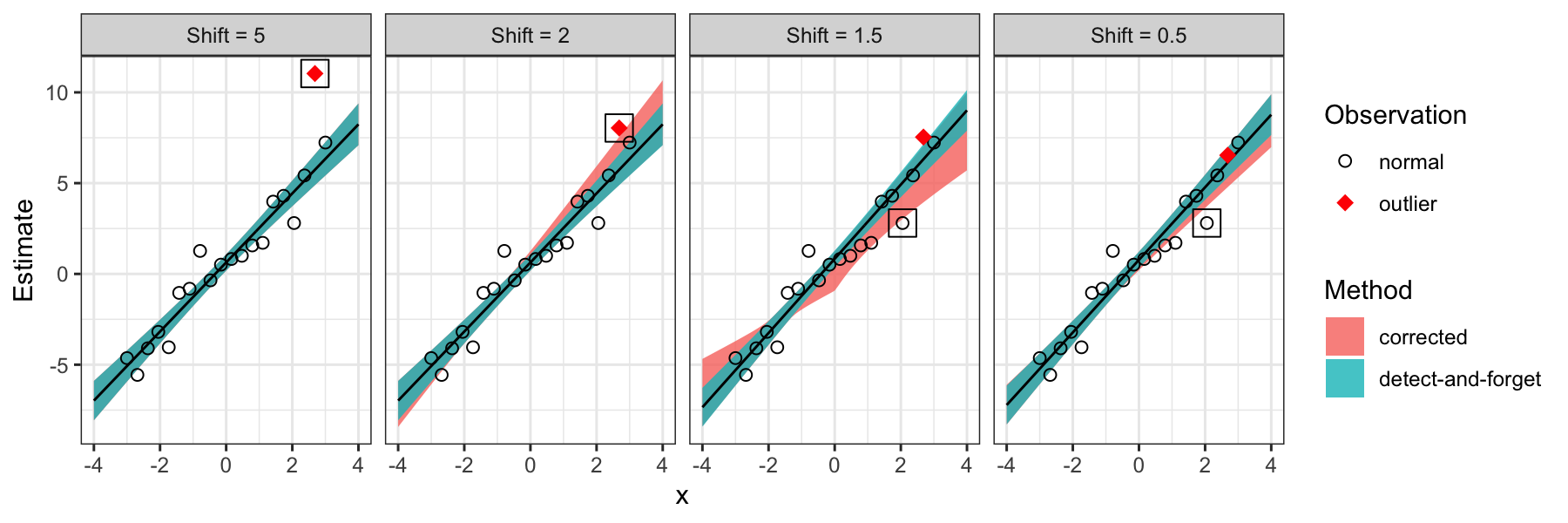}
\caption{Confidence intervals for the regression surface. Normal data are in black while the only outlier is marked in red. The point with an open black box is the detected outlier which has the largest Cook's distance. The black line is the regression line fitted using the data in which the detected outlier is removed.}
\label{fig:ci_surface_sequential_detection}
\end{figure}

We will now illustrate that even under symmetric errors, 
the \old~strategy can be problematic when performing inference for each covariate.
As a toy example, consider the situation shown in Figure \ref{fig:ci_surface_sequential_detection}, in which there are 19 ``normal'' points (in black), and a single ``outlier'' point (in red) has been shifted upward by different magnitudes.  For this illustration, we use a well-known approach for outlier detection called {\em Cook's distance} \citep{cook1977detection}:  
\begin{equation}
\label{eq:def_of_cook_distance}
  D_i = \frac{\widehat \varepsilon_i^2}{p \widehat \sigma^2} \frac{h_{ii}}{(1-h_{ii})^2},
\end{equation}
where $\widehat \varepsilon_i$ is the $i$-th residual from OLS on the entire data set, $\widehat \sigma^2:= \|\widehat \varepsilon\|_2^2/(n-p)$ is the scaled sum of squares, and $h_{ii}$ is the $i$-th diagonal entry of the hat matrix $X(X^TX)^{-1}X^T$. We declare the observation with the largest Cook's distance to be the outlier (indicated in the figure by an open black box) and then refit the regression model with this point removed (black regression line).  We then construct confidence intervals for the regression surface in two different ways: 
first, using the traditional \old~strategy, which ignores the outlier
removal step,  and second using \newknownsig, a method we will
introduce in this paper, which properly corrects for the removal. When
the outlier is obvious (leftmost panel), our method makes no
discernible correction.  With such a pronounced separation between the
outlier and non-outliers, Step 1 is unlikely to have removed a
non-outlier, and thus the distribution of the data for inference is
likely unaltered.   However, when the outlier is less easily
distinguished from the data, our corrected confidence intervals are
noticeably different from the classical ones.  
In particular, the
\newknownsig~intervals are pulled in the direction of the removed data point, thereby accounting for the possibility that the removed point may not in fact have been an outlier.

\begin{figure}[ht]
\centering
\includegraphics[width = \textwidth]{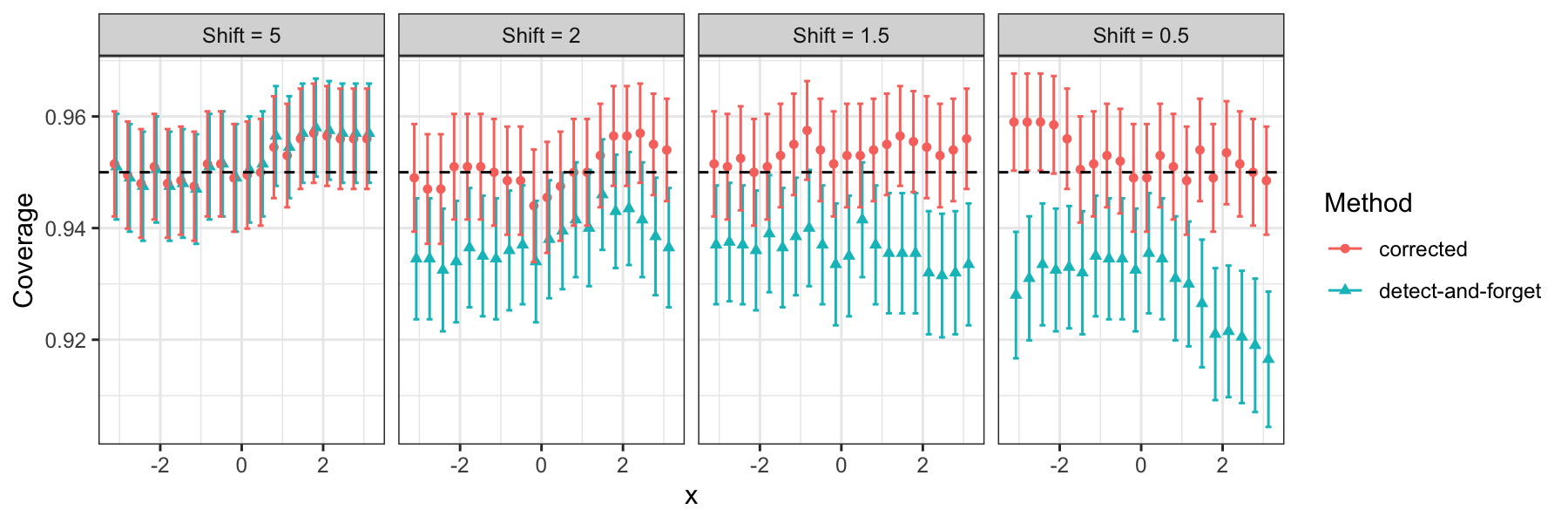}
\caption{Empirical coverage probability along the regression surface
(across $2000$ realizations). The dashed line represents $95\%$ coverage.}
\label{fig:cov_surface_sequential_detection}
\end{figure}

While Figure \ref{fig:ci_surface_sequential_detection} shows only a single realization of the two intervals in four different scenarios,
Figure \ref{fig:cov_surface_sequential_detection} shows the empirical coverage probability, averaged over 2000 realizations, of these two types of confidence intervals along the regression surface for the same four scenarios.
We see that when the outlier signal is strong (leftmost panel), both \old~and \newknownsig~intervals achieve $95\%$ coverage, as desired. However, as the outlier signal decreases, the \old~intervals begin to break down, while our \newknownsig~intervals remain unaffected. Indeed, we will show in this paper how all sorts of inferential statements (confidence intervals for regression coefficients, coefficient $t$-tests, $F$-tests, etc.) can be thrown off using a \old~strategy but can be corrected with a proper accounting for the outlier detection and removal step.


The machinery underlying our methodology is built on recent advances in \emph{selective inference} \citep{fithian2014optimal,taylor2015statistical}, specifically the framework introduced in \citet{lee2016exact,loftus2015selective}, and it fits within the framework of \emph{inferactive data analysis} introduced by \citet{2017arXiv170706692B}.
We give a brief introduction to the philosophy of selective inference in the context of outlier detection and refer readers to \citet{fithian2014optimal} for more details.

We assume a standard regression setting, $(\mathbf{y}, X) \in \mathbb{R}^n \times \mathbb{R}^{n \times p}$ with $\mathbf{y} \sim N(\boldsymbol{\mu}, \sigma^2 I_n)$, where $\boldsymbol{\mu}_i=\bx_i^T\boldsymbol{\beta}^*$ for $i\in M^*$ and $\bx_i$ is the $i$-th row of $X$.
Here $M^*$ is the set of {\em non-outliers}.
If $M$ is the set of detected {\em non-outliers}, then the \old~strategy forms the OLS estimator on the subset of observations in $M$, $\hat{\boldsymbol\beta}^M=X_{M, \bigcdot}^+\mathbf{y}_M$ (where $\mathbf{y}_M$ and $X_{M, \bigcdot}$ are formed by taking rows indexed by $M$, and $X_{M, \bigcdot}^+$ is the Moore-Penrose pseudoinverse of $X_{M, \bigcdot}$), and then proceeds with inference assuming that
$
\hat{\boldsymbol\beta}^M\sim N(\boldsymbol\beta^*, \sigma^2 [X_{M, \bigcdot}^TX_{M, \bigcdot}]^{-1}).
$

However, the above assumes that $M$ is non-random (or at least independent of $\mathbf{y}$) and that $M\subseteq M^*$, i.e. all true outliers have been successfully removed. 
However, in practice the set of declared non-outliers is in fact a function of the data, $\hM (\mathbf{y},X)$, and thus to perform inference would in principle require an understanding of the distribution of the much more complicated random variable
$
\hat{\boldsymbol\beta}^{\hM(\mathbf{y},X)}=X_{\hM(\mathbf{y},X), \bigcdot}^+\mathbf{y}_{\hM(\mathbf{y},X)}.
$

For general outlier removal procedures $\widehat M (\mathbf{y},X)$, such as ``make plots and inspect by eye'', the above distribution may be completely unobtainable. However, in this paper we define a class of outlier removal procedures for which the {\em conditional} distribution
$
\hat{\boldsymbol\beta}^{\hM(\mathbf{y},X)}~|~\hM(\mathbf{y},X)
$
can be precisely characterized.
Access to this conditional distribution will allow us to construct confidence intervals and $p$-values that are valid conditional on the set of outliers selected.

For example, we will produce a procedure for forming outlier-removal-aware confidence intervals $C_j^M(\mathbf{y},X)$ such that
$
\mathbb{P}(\boldsymbol{\beta}^*_j \in C^M_j(\mathbf{y},X) \ | \ \widehat M(\mathbf{y},X) = M) \geq 1 - \alpha 
$
for all subsets $M$ that do not include a true outlier.

If one could be certain that $\widehat M(\mathbf{y},X)\subseteq M^*$ (i.e., the procedure is adjusted to be sufficiently conservative and outliers are known to be sufficiently large), then such conditional coverage statements can be translated into a marginal (i.e., traditional) coverage statement:
$
\mathbb{P}(\boldsymbol{\beta}^*_j \in C^{\hM(\mathbf{y},X)}_j(\mathbf{y},X)) \geq 1 - \alpha.
$

However, in practice we do not know if all true outliers have been successfully removed. If $\widehat M(\mathbf{y},X)\not\subseteq M^*$, then OLS is no longer guaranteed to produce an unbiased estimate of $\boldsymbol{\beta}^*$.
OLS performed on the observations in $M$ instead estimates a parameter $\boldsymbol{\beta}^M$, which depends on both $\boldsymbol{\beta}^*$ and on $\mu_{M\setminus M^*}$, the mean of the true outliers that were not detected:
\begin{equation}
\label{eq:param_after_removal}
  \boldsymbol{\beta}^M := \underset{\boldsymbol{\beta} \in \mathbb{R}^p}{\arg \min} \  \mathbb{E}[\| \mathbf{y}_{M} - X_{M, \bigcdot} \boldsymbol{\beta} \|_2^2] = X_{M, \bigcdot}^+ \boldsymbol{\mu}_M.
\end{equation}

The goal of this paper is not to improve the performance of outlier removal procedures---certainly there is already extensive work in the literature on outlier removal. 
Rather, our goal is to provide valid inferential statements for someone who has chosen to use a particular outlier removal procedure, $\hM(\mathbf{y},X)$.
Thus, to stay within the scope of this problem, we will simply acknowledge that if a procedure $\hM(\mathbf{y},X)$ is prone to failing to identify outliers, then one cannot hope to estimate $\boldsymbol{\beta}^*$ but must instead focus on estimating and performing inference for $\boldsymbol{\beta}^{\hM(\mathbf{y},X)}$, which reflects more accurately than $\boldsymbol{\beta}^*$ the relationship between $X$ and $\mathbf{y}$ in the data that is provided to us by $\hM(\mathbf{y},X)$.
For example, we will provide intervals with guaranteed coverage of $\boldsymbol{\beta}^{\hM(\mathbf{y},X)}$:
$
\mathbb{P}(\boldsymbol{\beta}^{\hM(\mathbf{y},X)}_j \in C^{\hM(\mathbf{y},X)}_j(\mathbf{y},X)) \geq 1 - \alpha.
$
We will likewise provide all the standard confidence intervals and hypothesis tests for regression but focused on $\boldsymbol{\beta}^{\hM(\mathbf{y},X)}$ in place of $\boldsymbol{\beta}^*$.

This discussion emphasizes the inherently different effect of false positives (i.e., removing points that are not true outliers) versus false negatives (i.e., failing to remove points that are true outliers). When all true outliers are removed, $\boldsymbol{\beta}^{\hM} = \boldsymbol{\beta}^*$, and our machinery gives corrected inferential statements that account for the outlier removal step (including accounting for any false positives). By contrast, when true outliers remain, $\boldsymbol{\beta}^{\hM}\neq \boldsymbol{\beta}^*$, both \old~and our procedure give inferential statements about $\boldsymbol{\beta}^{\hM}$ rather than $\boldsymbol{\beta}^*$; however, in the case of our method, these statements are at least valid.

The rest of the paper is organized as follows: in Section \ref{sec:formulation} we formulate the problem more precisely and describe the class of outlier detection procedures over which our framework applies; Section \ref{sec:method} describes our methodology for forming confidence intervals and extracting $p$-values that are properly corrected for outlier removal; Section \ref{sec:example} provides empirical comparisons of the naive \old~strategy and our method, both through comprehensive simulations and a re-analysis of three real data sets; Section \ref{sec:discussion} gives a discussion and possible next steps. 
A companion {\tt R} package, {\tt outference}, is available at \url{https://github.com/shuxiaoc/outference}. 
For brevity, we collect proofs of most theoretical results, some additional simulation results and the implementation details in the online supplementary material. 

We conclude this section by introducing some notation that will be used throughout this paper. 
For $n \in \mathbb{N}$, we let $[n]:= \{1, 2, \hdots, n\}$. For a matrix $X$, we let $\mathscr{C}(X)$ be its column space and $\text{tr}(X)$ be its trace. 
We let $X_{I, J}$ be the submatrix formed by rows and columns indexed by $I$ and $J$, respectively, and we let $X_{I, \bigcdot}$ be the submatrix formed by rows indexed by $I$.
We let $P_X$ be the projection matrix onto $\mathscr{C}(X)$ and $P_X^\perp := I - P_X$.
For a submatrix $X_{I,J}$, we write $P_{I, J} := P_{X_{I, J}}$ when there is no ambiguity.
We use $\indep$ to denote statistical independence.

\section{Problem Formulation}\label{sec:formulation}

\subsection{The General Setup}
We elaborate on the framework described in the previous section, introducing some additional notation. 
We assume $\mathbf{y} = \boldsymbol{\mu} + \boldsymbol{\varepsilon}$, where $\mathbf{y} \in \mathbb{R}^n, \boldsymbol{\varepsilon} \sim N(\mathbf{0}, \sigma^2 I_n)$ and consider the \emph{mean-shift model},
\begin{equation}
\label{eq:mean_shift_model}
  \boldsymbol{\mu} = X \boldsymbol{\beta}^* + \mathbf{u}^*,
\end{equation}
where $\boldsymbol{\beta}^* \in \mathbb{R}^p$, and $X$ is a non-random matrix of predictors.
The set $M^* = [n]\setminus \text{supp}(\mathbf{u}^*)$ is the index set of \emph{true non-outliers}; equivalently, $M^{*c}$ is the index set of true outliers.
By definition of $M^*$, $\mathbf{u}^*_{M^*} = \mathbf{0}$ and $\mathbf{u}^*_{i}\neq0$ for $i\in M^{*c}$.
This setup assumes that all outliers considered are ``vertical'' in the sense that they only contaminate the model in the $y$-direction.
We denote a data-dependent outlier removal procedure, $\widehat M: \mathbb{R}^n \to 2^{[n]}$, as a function mapping the data $\mathbf{y}$ to the index set of \emph{detected non-outliers} (for notational ease, we suppress the dependence of $\hM$ on $X$ since $X$ is treated as non-random). 
We will assume throughout that $X_{\hM(\mathbf{y}), \bigcdot}$ has linearly independent columns.

For a fixed subset of the observations $M\subseteq [n]$, the parameter $X\boldsymbol{\beta}^M$ where $\bbeta^M$ is defined in \eqref{eq:param_after_removal} represents the \emph{best linear approximation} of $\boldsymbol{\mu}_M$ using the $p$ predictors in $X_{M,\bigcdot}$.
In what follows, we will provide hypothesis tests and confidence intervals for $\boldsymbol{\beta}^{M}$ conditional on the event $\{\hM(\mathbf{y})=M\}$.

Combining \eqref{eq:param_after_removal} and \eqref{eq:mean_shift_model} with the assumption that $X_{M,\bigcdot}$ has linearly independent columns, we have
$  \boldsymbol{\beta}^M= X_{M, \bigcdot}^+ (X_{M,\bigcdot} \boldsymbol{\beta}^* + \mathbf{u}^*_M)
= \boldsymbol{\beta}^*
  + X_{M, \bigcdot}^+ \mathbf{u}^*_M $.
Since $\mathbf{u}^*_{M^*}=0$, it follows that $\boldsymbol{\beta}^M=\boldsymbol{\beta}^*$ when $M\subseteq M^*$.
This result makes it clear that if one wishes to make statements about $\beta^*$, then one must ensure that the procedure $\hM$ is screening out all outliers. 

Our focus will be on performing inference on $\boldsymbol{\beta}^M$ conditional on the event $\{\hM(\mathbf{y})=M\}$.
Importantly, such inferential procedures in fact provide asymptotically valid inferences for $\boldsymbol{\beta}^*$ as long as one's outlier removal procedure asympotically detects all outliers.
For example, the next proposition establishes that confidence intervals providing conditional coverage of $\boldsymbol{\beta}^M_j$ given $\{\hM(\mathbf{y})=M\}$ do in fact achieve traditional (i.e., unconditional) coverage of $\boldsymbol{\beta}^*_j$ asymptotically if one is using an outlier detection procedure that is guaranteed to screen out all outliers as $n\to \infty$.

{\prop
\label{prop:unconditional_coverage}
For $\mathbf{y}$ generated through the mean-shift model \eqref{eq:mean_shift_model}, consider intervals $C_j^{\hM}$, satisfying $\mathbb{P}(\boldsymbol{\beta}^{\hM}_j \in  C_j^{\widehat M} \ | \ \widehat M = M) = 1-\alpha$. If the outlier detection procedure $\hM$ satisfies $\mathbb{P}(\widehat M \subseteq M^*) \to 1$ as $n \to \infty$, then we have $\mathbb{P}(\boldsymbol{\beta}^*_j \in C_j^{\widehat M}) \to 1-\alpha$.
}

This proposition is based on two simple observations: first, that conditional coverage of $\boldsymbol{\beta}^{\hM}$ implies unconditional coverage of $\boldsymbol{\beta}^{\hM}$; second, that $\widehat M \subseteq M^*$ implies that $\boldsymbol{\beta}^{\widehat M} = \boldsymbol{\beta}^*$.

Such a screening property is reasonable to demand of an outlier detection procedure, and related results exist in the literature \citep{zhao2013high}.
For example, consider using Cook's distance \eqref{eq:def_of_cook_distance} to detect outliers:
\begin{equation}
\label{eq:cook_out_det}
  \widehat M(\mathbf{y}) = \{i: D_i < \lambda/n\},
\end{equation}
where $\lambda$ is a prespecified cutoff.
In Section 2 of the supplementary material, we provide conditions (based on a result of \citealt{zhao2013high}) under which $\mathbb{P}(\hM =M^*) \to 1$ for an appropriate choice of $\lambda$.
While $\lambda=0$ would trivially satisfy the screening property, we of course need a procedure that leaves sufficient observations for estimation and inference.

While the mean-shift model model is common in the outlier detection literature, it is by no means the only reasonable one (see, e.g., \citealt{huber1965robust,thompson1985note,huber1992robust}). We choose to focus on the mean-shift model because it provides a simple yet practical working definition of an ``outlier'', and it is relatively easy to prove the effectiveness of the outlier detection procedure considered in this paper under this model (e.g., Proposition S.1 in the supplementary material). However, our main results are, indeed, independent of the choice of specific outlier model. 

\subsection{Quadratic Outlier Detection Procedures}\label{sec:quad}
In this section we define a general class of outlier detection procedures for which our methodology will apply. 
We then show that this class includes several of the most famous outlier detection procedures.

{\definition \label{def:quad_out_det}
We say an outlier detection procedure is \emph{quadratic} if the event
${\{ \widehat M(\mathbf{y}) = M \}}=:\mathcal{E}_M$ is of the form
$\mathfrak{X}( \{\mathcal{E}_{M, i} \}_{i \in I_M})$, where
$\mathfrak{X}$ denotes a general set operator that maps a finite
family of sets to a single set, $I_M$ is a finite index set, and $\mathcal{E}_{M, i} := \{ \mathbf{y} \in \mathbb{R}^n: \mathbf{y}^T Q_{M, i} \mathbf{y} + \mathbf{a}_{M, i}^T \mathbf{y} + b_{M, i} \geq 0 \}$,
for some $Q_{M, i} \in \mathbb{R}^{n \times n}, \mathbf{a}_{M, i} \in
\mathbb{R}^n$, and $b_{M, i} \in \mathbb{R}$.}

Generally, $\mathfrak{X}$ should be thought of as taking finite unions, intersections, and complements. The above definition is a direct generalization of Definition 1.1 of \citet{loftus2015selective}, in which $\mathfrak{X} \equiv \bigcap$. 
We will see that many outlier detection procedures are quadratic in the sense of Definition \ref{def:quad_out_det}.
While most of the time the definition in \citet{loftus2015selective} will apply, there are certain cases that require our generalization (see Section 3.2 of the supplementary material for a specific example).
The next proposition shows that outlier detection using Cook's distance is quadratic in the sense of Definition \ref{def:quad_out_det}. 

{\prop
\label{prop:cook_is_quadratic}
Outlier detection using Cook's distance \eqref{eq:cook_out_det} is \emph{quadratic} with
\begin{align}
  \label{eq:cook_out_det_intersection} \mathcal{E}_M &= \bigcap_{i \in [n]} \mathcal{E}_{M, i}, \\
  \label{eq:cook_out_det_each_slice} \mathcal{E}_{M, i} &=
  \bigg\{\mathbf{y} \in \mathbb{R}^n: (-1)^{\mathbbm{1}\{i\in M^c\}} \mathbf{y}^T \bigg(\frac{\lambda p}{n}(1-h_i)^2 P_X^\perp - (n-p)h_i P_X^\perp \mathbf{e}_i \mathbf{e}_i^T P_X^\perp \bigg) \mathbf{y} > 0  \bigg\},
\end{align}
where $\mathbf{e}_i$ is $i$-th standard basis for $\mathbb{R}^n$ and $h_i = (P_X)_{ii}$.}
{\proof We may write $D_i = \bigg( \mathbf{y}^T (P_X^\perp e_i e_i^T P_X^\perp )\mathbf{y} \bigg/  \mathbf{y}^T P_X^\perp \mathbf{y}  \bigg) \cdot \bigg((n-p)h_i \bigg / p(1 - h_i)^2\bigg).$ Plugging this expression to $\widehat M = \{i : D_i < \lambda/n\}$ and $\widehat M^c = \{i: D_i \geq \lambda/n\}$ gives the desired result. \qed}

As a second example, we consider Huber's M-estimator \eqref{eq:huber}. Though it is a robust regression method, (as observed in \citealt{she2011outlier}) its solution $\hat \beta_\lambda$ can be equivalently expressed as the following lasso program \citep{tibshirani1996regression} within the context of the mean-shift model:
\begin{align}
\label{eq:ipod}
  (\hat{\boldsymbol{\beta}}_\lambda, \hat{\mathbf{u}}_\lambda) = \underset{\boldsymbol{\beta} \in \mathbb{R}^p, \mathbf{u}
    \in \mathbb{R}^n}{\arg \min} \ \frac{1}{2n} \| \mathbf{y} - X \boldsymbol{\beta} - \mathbf{u}
  \|_2^2 + \lambda \| \mathbf{u} \|_1,
\end{align}
which the authors refer to as the {\em soft-IPOD} method.
The $\ell_1$-penalty induces sparsity in $\hat{\mathbf{u}}_\lambda$, and one takes
$\hM(\mathbf{y}) = \{ i: \hat{\mathbf{u}}_{\lambda,i} = 0 \}$ as the detected non-outliers. The outliers correspond to the elements whose residuals are in the quadratic (rather than linear) region of Huber's loss function. In Section 3 of the supplementary material, this approach is shown to be a quadratic outlier detection procedure, which explains why our framework can accommodate this foundational robust regression method.
The DFFITS outlier detection method \citep{welsch1977linear} is described in Section 3 of the supplementary material, where it is shown to be quadratic.
Extending our framework to state-of-the-art outlier detection methods (e.g., examining the residuals after MM-estimation) remains an open question.

\section{Inference Corrected for Outlier Removal}\label{sec:method}
In this section, we describe how the standard inferential tools of OLS can be corrected to account for outlier removal. 
The only requirement is that the outlier detection procedure be quadratic (as defined in the previous section).
The inferential statements are made conditional on the event $\{\hM(\mathbf{y})=M\}$ and are about the parameter $\boldsymbol{\beta}^M$.
As previously discussed, such statements translate to unconditional statements about $\beta^*$ when $\hM\subseteq M^*$, that is, when all true outliers are removed.
Section \ref{subsec:sigma-known} treats the case in which $\sigma$ is known.
Section \ref{subsec:sel_test_unknown_sig} provides procedures for the case when $\sigma$ is unknown.

\subsection{Confidence Intervals and Hypothesis Tests When \texorpdfstring{$\sigma$}{G} Is Known}\label{subsec:sigma-known}
In this section, we suppose that $\sigma$ is known and provide confidence intervals and hypothesis tests.
In the classical setting, inference is based on the normal and $\chi^2$ distributions and typically involves individual regression coefficients $\boldsymbol{\beta}^M_j$, the regression surface $\boldsymbol{x}_0^T \boldsymbol{\beta}^M$, or groups of regression coefficients $\boldsymbol{\beta}_g^M$.
We begin by observing that both $\boldsymbol{\beta}^M_j$ and $\mathbf{x}_0^T \boldsymbol{\beta}^M$ are of the form $\boldsymbol{\nu}^T \boldsymbol{\mu}$ for some vector $\boldsymbol{\nu}$ that depends on $M$: $\boldsymbol{\beta}^M_j = \mathbf{e}_j^T X_{M, \bigcdot}^+ I_{M, \bigcdot} \boldsymbol{\mu}$ and $\mathbf{x}_0^T \boldsymbol{\beta}^M = \mathbf{x}_0^T X_{M, \bigcdot}^+ I_{M, \bigcdot} \boldsymbol{\mu}$.
The next theorem gives a unified treatment of these two cases that will allow us to construct confidence intervals and $p$-values that properly account for outlier removal.  
{\theorem
\label{thm:pw_inf_lincomb_mu}
Assume the outlier detection procedure $\{\widehat M = M\}$ is quadratic as in Definition \ref{def:quad_out_det}. Let $\boldsymbol{\nu}\in\mathbb{R}^n$ be a vector that may depend on $M$. Define
\begin{align*}
  \mathcal{Z} & := \frac{\boldsymbol{\nu}^T \mathbf{y}}{\sigma \|\boldsymbol{\nu}\|_2}, \ \ \ \mathbf{z} := P_{\boldsymbol{\nu}}^\perp \mathbf{y} = \bigg(I - \frac{\boldsymbol{\nu} \boldsymbol{\nu}^T}{\| \boldsymbol{\nu} \|_2^2} \bigg) \mathbf{y}.
\end{align*}
We have
  \begin{equation}
  \label{eq:Z_trunc_norm}
    \mathcal{Z} \ \bigg| \  \bigg\{ \widehat M = M, \mathbf{z}  \bigg\} \sim TN\bigg(\frac{\boldsymbol{\nu}^T \boldsymbol{\mu}}{\sigma \|\boldsymbol{\nu}\|_2} , 1  ;  E_{M, \mathbf{z}}\bigg),
  \end{equation}
  where the R.H.S is a $N(\frac{\boldsymbol{\nu}^T \boldsymbol{\mu}}{\sigma \|\boldsymbol{\nu}\|_2} , 1)$ random variable truncated to the set $E_{M, \mathbf{z}}$. The truncation set is defined in Section 4.2 of the supplementary material and can be computed by finding the roots of a finite set of quadratic polynomials.
  Thus, letting $F_{\xi, \gamma^2}^{E}$ be the CDF of a $TN(\xi, \gamma^2; E)$ random variable, we have
  \begin{equation}
  \label{eq:p_val_beta_j}
    1 - F^{E_{M, \bz}}_{\frac{\bnu^T \bmu}{\sigma \|\bnu\|_2}, 1 }(\mathcal{Z}) \ \bigg| \ \{ \widehat M = M  \} \sim \text{\emph{unif}}(0, 1).
  \end{equation}}

The classical analogue to the above theorem is the (much simpler!) statement that $\mathcal{Z}\sim N(\bnu^T\bmu/[\sigma\|\bnu\|_2],1)$.

This theorem is essentially a generalization of \citet[Theorem 5.2]{lee2016exact} and a special case of \citet[Theorem 3.1]{loftus2015selective}; however, a key difference is that these works are focused on accounting for variable selection rather than outlier removal (which, in essence, is ``observation selection''). 

\subsubsection{Corrected Confidence Intervals}\label{subsec:sel_ci}
We begin by applying Theorem \ref{thm:pw_inf_lincomb_mu} to get confidence intervals corrected for outlier removal.
{\cor 
\label{cor:pw_inf_lincomb_mu_ci}
Under the conditions and notation of Theorem~\ref{thm:pw_inf_lincomb_mu},
if we find $L$ and $U$ such that
\begin{align}
  \label{eq:pw_lincomb_mu_lo_and_up} L: F_{\frac{L}{\sigma \|\bnu\|_2}, 1}^{E_{M, \bz}}(\mathcal{Z}) = 1 - \frac{\alpha}{2} , \ \ \
   U: F_{\frac{U}{\sigma \|\bnu\|_2}, 1}^{E_{M, \bz}}(\mathcal{Z}) = \frac{\alpha}{2},
\end{align}
then $[L, U]$ is a valid $(1-\alpha)$ selective confidence interval for $\bnu^T \bmu$.  That is,
\begin{equation}
\label{eq:pw_coverage_for_beta_j}
  \mathbb{P}(\bnu^T \bmu \in [L, U] \ | \ \widehat M = M) = 1-\alpha.
\end{equation}}

This result encompasses the two most common types of confidence intervals arising in regression: intervals for the regression coefficients $\bbeta^M_j$ and intervals for the regression surface $\bx_0^T \bbeta^M$.

{\cor 
\label{cor:pw_inf_coef_and_surface}
We write $\bbeta^M_j = \bnu_{\text{\emph{coef}}, j}^T \bmu$ and $\bx_0^T \bbeta^M = \bnu_{\text{\emph{surf}}}^T \bmu$, where $\bnu_{\text{\emph{coef}}, j} = (\be_j^T X_{M, \bigcdot}^+ I_{M, \bigcdot})^T$ and $\bnu_{\text{\emph{surf}}} = (\bx_0^T X_{M, \bigcdot}^+ I_{M, \bigcdot})^T$. Then Theorem \ref{thm:pw_inf_lincomb_mu} and Corollary \ref{cor:pw_inf_lincomb_mu_ci} apply.}

A third type of interval common in regression is the prediction interval, intended to cover $\bx_0^T \bbeta^M + \bvarepsilon_0$, where $\bx_0\in\mathbb{R}^p$ is a new data point and $\bvarepsilon_0 \sim N(0, \sigma^2)$ is independent of $\bvarepsilon$.
While $\bnu^T \by \ | \ \{\widehat M = M, \bz\}$ is a truncated normal random variable, $\bnu^T \by + \bvarepsilon_0 \ | \ \{\widehat M = M, \bz \}$ is not, so the strategy adopted in Theorem \ref{thm:pw_inf_lincomb_mu} does not directly apply to this case. Instead we employ a simple (but conservative) strategy.

{\prop
\label{prop:pw_prediction_int}
Let $\varepsilon_0 \sim N(0, \sigma^2)$ be the noise independent of $\by$. For a given significance level $\alpha \in (0, 1)$, let $\widetilde \alpha \in (0, \alpha)$. Given $\bx_0 \in \mathbb{R}^p$, let $[L_{\widetilde \alpha}, U_{\widetilde \alpha}]$ be the $(1-\widetilde \alpha)$ selective confidence intervals for $\bx_0^T \bbeta^M$ as defined in \eqref{eq:pw_lincomb_mu_lo_and_up}. Then we have
\begin{equation}
\label{eq:pw_prediction_int}
  \mathbb{P}\bigg(L_{\widetilde \alpha} -  \Phi^{-1}(1-\frac{\alpha - \widetilde \alpha}{2})\sigma \leq \bx_0^T \bbeta^M + \varepsilon_0 \leq U_{\widetilde \alpha} +  \Phi^{-1}(1-\frac{\alpha - \widetilde \alpha}{2})\sigma \ \bigg| \ \widehat M = M\bigg) \geq 1-\alpha,
\end{equation}
  where $\Phi$ is the CDF of a standard normal distribution.
}

In practice, we can optimize over $\widetilde \alpha$ so that the length of the interval is minimized.

\subsubsection{Corrected Hypothesis Tests}\label{subsec:sel_test_know_sig}
Theorem \ref{thm:pw_inf_lincomb_mu} allows us to form selective hypothesis tests about the parameter $\bnu^T\bmu$ where $\bnu$ may depend on the selected index set of observations $M$.  
{\cor
\label{cor:pw_inf_lincomb_mu_pvalue}
Under the conditions and notation of Theorem \ref{thm:pw_inf_lincomb_mu}, the quantity $1 - F^{E_{M, \bz}}_{0, 1 }(\mathcal{Z})$
gives a valid selective $p$-value for testing $H_0: \nu^T \mu = 0$.}

The most common application of the above would be for testing whether a specific regression coefficient is zero, conditional on $M$ being the selected set of non-outliers: $H_0(M, j): \bbeta^M_j = 0$ for $j \in [p]$.

As a generalization, we next focus on testing $H_0(M, g): \bbeta^M_g = 0$ for $g \subseteq [p]$. We begin with an alternative characterization of $H_0(M, g)$.

{\prop
\label{prop:test_nested_models_as_proj}
Set $\widetilde X_{M, g} = (I_{|M|} - P_{M, g^c})X_{M, g}$, where $P_{M, g^c}$ is the projection matrix onto $\mathscr{C}(X_{M, g^c})$. Let $\widetilde P_{M, g}$ be the projection matrix onto $\mathscr{C}(\widetilde X_{M, g})$. Then we have 
\begin{equation}
\label{eq:equiv_null_hypothesis_tiledP}
  \bbeta^M_g = \b0 \Leftrightarrow \widetilde X_{M, g}^+ \bmu_M = \b0 \Leftrightarrow \widetilde P_{M, g} \bmu_M = \b0.
\end{equation}
Further, define
$
  \widecheck P_{M, g}: =
  \begin{pmatrix}
    \widetilde P_{M, g} & \b0_{|M| \times (n-|M|)} \\
    \b0_{(n-|M|) \times |M|} & \b0_{(n-|M|)\times(n-|M|)}
  \end{pmatrix}.
$
Then $\widecheck P_{M, g}$ is an orthogonal projection matrix (it is symmetric and idempotent), and we have
\begin{equation}
\label{eq:equiv_null_hypothesis_checkP}
\bbeta^M_g = \b0 \Leftrightarrow \widecheck P_{M, g} \bmu = \b0. 
\end{equation}}

This proposition characterizes $H_0(M, g)$ as testing the projection of $\bmu$. In the non-selective case, testing $P\bmu = \b0$ for some projection matrix $P$ can be done based on $\sigma^{-2}\by^T P \by \sim \chi^2_{\text{{tr}}(P)}$ under $P\bmu = \b0$. We would expect that in the selective case, such tests can be done based on a truncated $\chi^2$ distribution.

{\theorem
\label{thm:test_nested_models}
Assume the outlier detection procedure $\{\widehat M = M\}$ is quadratic as in Definition \ref{def:quad_out_det}. Define
\begin{align}
  \label{eq:test_statistic_nested_models}\mathcal{X} &:= \frac{\|\widecheck P_{M, g} \by \|_2}{\sigma}, \ \ \  
  \bw := \frac{\widecheck P_{M, g} \by}{\| \widecheck P_{M, g} \by \|_2} = \frac{\widecheck P_{M, g} \by}{\sigma \mathcal{X}}, \ \ \
  \bz := \widecheck P_{M, g}^\perp \by.
\end{align}
Under $H_0(M, g): \bbeta^M_g = 0$, we have
\begin{equation}
\label{eq:trunc_chi2_nested_models}
  \mathcal{X}^2 \ \bigg| \ \{ \widehat M = M, \bw, \bz \} \sim T\chi^2_{\text{\emph{tr}}(\widecheck P_{M, g})} ( E_{M, \bw, \bz}) ,
\end{equation}
where the R.H.S is a central $\chi^2$ random variable with $\text{df} = \text{\emph{tr}}(\widecheck P_{M, g})$ truncated to the set $E_{M, \bw, \bz}$. The truncation set is defined in Section 4.5 of the supplementary material and can be computed by finding the roots of a finite set of quadratic polynomials.
Further, letting $F_{df}^{E}$ be the CDF of a $T\chi^2_{df}(E)$ random variable, we have
\begin{equation}
\label{eq:p_value_nested_models}
1 - F_{\text{\emph{tr}}(\widecheck P_{M ,g})}^{E_{M, \bw, \bz}}(\mathcal{X}^2) \ \bigg| \ \{ \widehat M = M \} \sim \text{\emph{unif}}(0, 1),
\end{equation}
which is a valid selective $p$-value for testing $H_0(M, g): (\bbeta^M)_g = 0$.}

This theorem is adapted from \citet[Theorem 3.1]{loftus2015selective} to the outlier detection context. In the special case where $g = j$ is a single index, direct computation can show that $\widecheck P_{M, j} = P_{\bnu_{\text{coef}, j}}$, so that
$
\mathcal{X}^2 = (\bnu^T_{\text{coef}, j} \by)^2/({\sigma \| \bnu_{\text{coef}, j} \|_2})^2,  w = \text{sign}(\bnu_{\text{coef}, j}^T \by) \bnu_{\text{coef}, j} $, and  $ \bz = P_{\bnu_{\text{coef}, j}}^\perp \by.$
Then this theorem nearly reduces to Theorem \ref{thm:pw_inf_lincomb_mu}, except that in this theorem, we need to condition on the sign of $\bnu_{\text{coef}, j}^T \by$.

\subsection{Extension to \texorpdfstring{$\boldsymbol{\sigma}$}{G} Unknown Case}\label{subsec:sel_test_unknown_sig}
In this section, we extend results in Section \ref{subsec:sel_test_know_sig} to the $\sigma$ unknown case. In the non-selective case, the hypothesis $H_0: \bbeta^*_g = \b0$ is equivalent to $H_0: \bmu \in \mathscr{C}(X_{\bigcdot, g^c})$.
Hence under whichever $H_0$, $(P_{\bigcdot, g^c}^\perp - P_X^\perp) \by$ and $P_X^\perp \by$ will both be centered normal random variables, and the test can be done based on $\mathcal{F} = \bigg((\|P_{\bigcdot, g^c}^\perp \by \|_2^2 - \|P_X^\perp \by \|_2^2)/|g| \bigg) \bigg/ \bigg({\| P_X^\perp \by \|_2^2 / (n-p)} \bigg) \sim F_{|g|, n-p}$.
By analogy, we might expect $H_0(M, g): \bbeta^M_g = \b0$ to be equivalent to $H_0: \bmu_M \in \mathscr{C}(X_{M, g^c})$, which would suggest that the test should be done based on a truncated $F$ distribution; however, we will see in the rest of the section that this is only partially true.

{\prop
\label{prop:two_nulls_not_equiv_F_test}
We have $\bmu_M \in \mathscr{C}(X_{M, g^c}) \Rightarrow \bbeta^M_g = \b0$ but $\bbeta^M_g = \b0 \nRightarrow \bmu_M \in \mathscr{C}({X_{M, g^c}})$. Moreover, if $M \subseteq M^*$, then $\bbeta^M_g = \b0 \Rightarrow \bmu_M \in \mathscr{C}({X_{M, g^c}})$.}

In order to form an $F$ statistic, we need both the numerator and the denominator to be composed of \emph{centered} random variables.
So it is necessary to assume $\bmu_M \in \mathscr{C}(X_{M, g^c})$. Hence this proposition says that testing $H_0: \bmu_M \in \mathscr{C}(X_{M, g^c})$ is the best we can do. 
Our next result adapts a truncated $F$ significance test from \citet{loftus2015selective} to our purposes.

{\theorem
\label{thm:sel_F_test}
Assume the outlier detection procedure $\{\widehat M = M\}$ is quadratic as in Definition \ref{def:quad_out_det}. Let $\bR_1 := P_\text{\emph{sub}}^\perp \by, \bR_2 := P_\text{\emph{full}}^\perp \by$, where
$
  P_\text{\emph{sub}} := 
  \begin{pmatrix} 
    P_{M, g^c} & \b0_{|M| \times (n - |M|)} \\
    \b0_{(n-|M|) \times |M|} & I_{(n-|M|)}
  \end{pmatrix}
$   
and
$
  P_\text{\emph{full}} := 
  \begin{pmatrix}
    P_{M, \bigcdot} & \b0_{|M| \times (n - |M|)} \\
    \b0_{(n - |M|) \times |M|}  & I_{(n - |M|)}
  \end{pmatrix}.
$  
Define
\begin{align}
  \label{eq:sel_F_test_def_of_F} \mathcal{F} &:=  \frac{(\| \bR_1\|_2^2 - \| \bR_2\|_2^2)/|g|}{\|\bR_2\|_2^2/(|M| - p)}, \\
  \label{eq:sel_F_test_def_of_w_and_z_and_r} \bw_\Delta & := \frac{\bR_1 - \bR_2}{\| \bR_1 - \bR_2 \|_2}, \ \ \ \bw_2 := \frac{\bR_2}{\|\bR_2\|_2}, \ \ \ \bz := P_\text{\emph{sub}} \by, \ \ \ r := \|\bR_1\|_2, \\
  \label{eq:sel_F_test_def_of_g1_and_g2} g_1(\mathcal{F}) & := \sqrt{\frac{|g| \mathcal{F}/(|M| - p)}{1 + |g|\mathcal{F}/(|M| - p)}}, \ \ \ g_2(\mathcal{F}):= \sqrt{\frac{1}{1 + |g|\mathcal{F}/(|M| - p)}}.
\end{align}
Under $H_0: \bmu_M \in \mathscr{C}(X_{M, g^c})$, we have
\begin{equation}
  \label{eq:sel_F_test_trunc_F} \mathcal{F} \ \bigg| \ \{ \widehat M = M, \bw_\Delta, \bw_2, \bz, r \} \sim TF_{|g|, |M| - p}(E_{M, \bw_\Delta, \bw_2, \bz, r}),
\end{equation}
where the R.H.S. is a central $F$ random variable with $df_1 = |g|, df_2 = |M| - p$ truncated to the set $E_{M, \bw_\Delta, \bw_2, \bz, r}$. The truncation set is defined in Section 4.7 of the supplementary material. 
Further, letting $F_{df_1, df_2}^E$ be the CDF of a $TF_{df_1, df_2}(E)$ random variable, we have
\begin{equation}
  \label{eq:sel_F_test_p_val}
  1 - F^{E_{M, \bw_\Delta, \bw_2, \bz, r}}_{|g|, |M| - p}(\mathcal{F}) \ \bigg| \ \{ \widehat M = M \} \sim \text{\emph{unif}}(0, 1),
\end{equation}
which is a valid selective $p$-value for testing $H_0: \bmu_M \in \mathscr{C}(X_{M, g^c})$.}

Computing the truncation set in the $\sigma$ unknown case is non-trivial since each slice is no longer a quadratic function in $\mathcal{F}$.
We adopt the strategy suggested by \citet[Section 4.1]{loftus2015selective}.
For completeness, we provide the details of their strategy (adapted to our notation) in the online supplementary material.

We conclude this section by noting that Theorem \ref{thm:sel_F_test} does not give us a way to construct confidence intervals for $\bbeta^M_j$.
In order to form confidence intervals for $\bbeta^M_j$, one would need to be able to test for $H_0: \bbeta^M_j = c_0$ for some non-zero constant $c_0$. 
Under this null, $\mathcal{F}$ does not necessarily reduce to the square of a truncated $t$ distribution: First, $\bmu_M \in \mathscr{C}(X_{M, \bigcdot})$ does not necessarily hold, and as a result, $\bR_2$ may not even be centered; second, the independence between $\mathcal{F}$ and $(\bw_\Delta, \bw_2, \bz, r)$ may not hold.
Hence the construction of confidence intervals does not follow directly from Theorem \ref{thm:sel_F_test} and is left as future work.

\section{Empirical Examples}\label{sec:example}

We provide simulations and real data examples in this section. We notice that our method requires evaluation of survival functions (equivalently, the CDFs) of truncated normal, $\chi^2$, $t$ and $F$ distributions. Our implementations are greatly inspired by that of {\tt selectiveInference} package \citep{selectiveInfPacakge}. We refer readers to the online supplementary material for more details.

\subsection{Simulations}
In this section, we focus on the case where the outlier detection is done by Cook's distance, and we assume $\sigma$ is unknown. We refer the readers to the supplementary materials for more detailed and comprehensive simulations. We compare the performance of the following three inferential procedures:
\setlist{nolistsep}
\begin{itemize}
  \item \old: After outlier detection, refit an OLS regression model using the remaining data $(\by_M , X_{M, \bigcdot})$ and do inference based on the classical (non-selective) theory (we use $t$ and $F$ distributions since $\sigma$ is unknown);
  \item \newest: Do selective inference as developed in Section \ref{subsec:sel_ci} and \ref{subsec:sel_test_know_sig}, with \emph{estimated} $\sigma$, and the estimation of $\sigma$ is done by
  $
    \hat \sigma^2_\text{EST} = \frac{1}{n - |S_\text{AUG}|} \| \by - X_\text{AUG} \hat \bbeta_\text{AUG} \|_2^2,
  $
  where we fit a lasso regression of $\by$ on $X_\text{AUG}= (X : I_n)$ to get $\hat \bbeta_\text{AUG} \in\mathbb R^{p+n}$, and $S_\text{AUG}$ is the support of $\hat \bbeta_\text{AUG}$.  \citet{reid2013study} demonstrate that such a strategy gives a reasonably good estimate of $\sigma^2$ in a wide range of situations.
  \item \newexact: Do selective inference assuming unknown $\sigma$ as developed in Section \ref{subsec:sel_test_unknown_sig} (note: this method does not give confidence intervals).
\end{itemize}
We fix $n = 100, p = 11$. 
Our indexing of variables starts from $0$ (i.e. $\bbeta^*_0$ corresponds to the intercept). 
The first column of $X$ is set to be $\mathbf{1}$ and the rest of the columns are generated from i.i.d. $N(0, 1)$ and scaled to have $\ell_2$ norm $\sqrt{n}$. 
We fix $\sigma = 1$.

\begin{figure}[ht]
\centering
\makebox{\includegraphics[width = \textwidth]{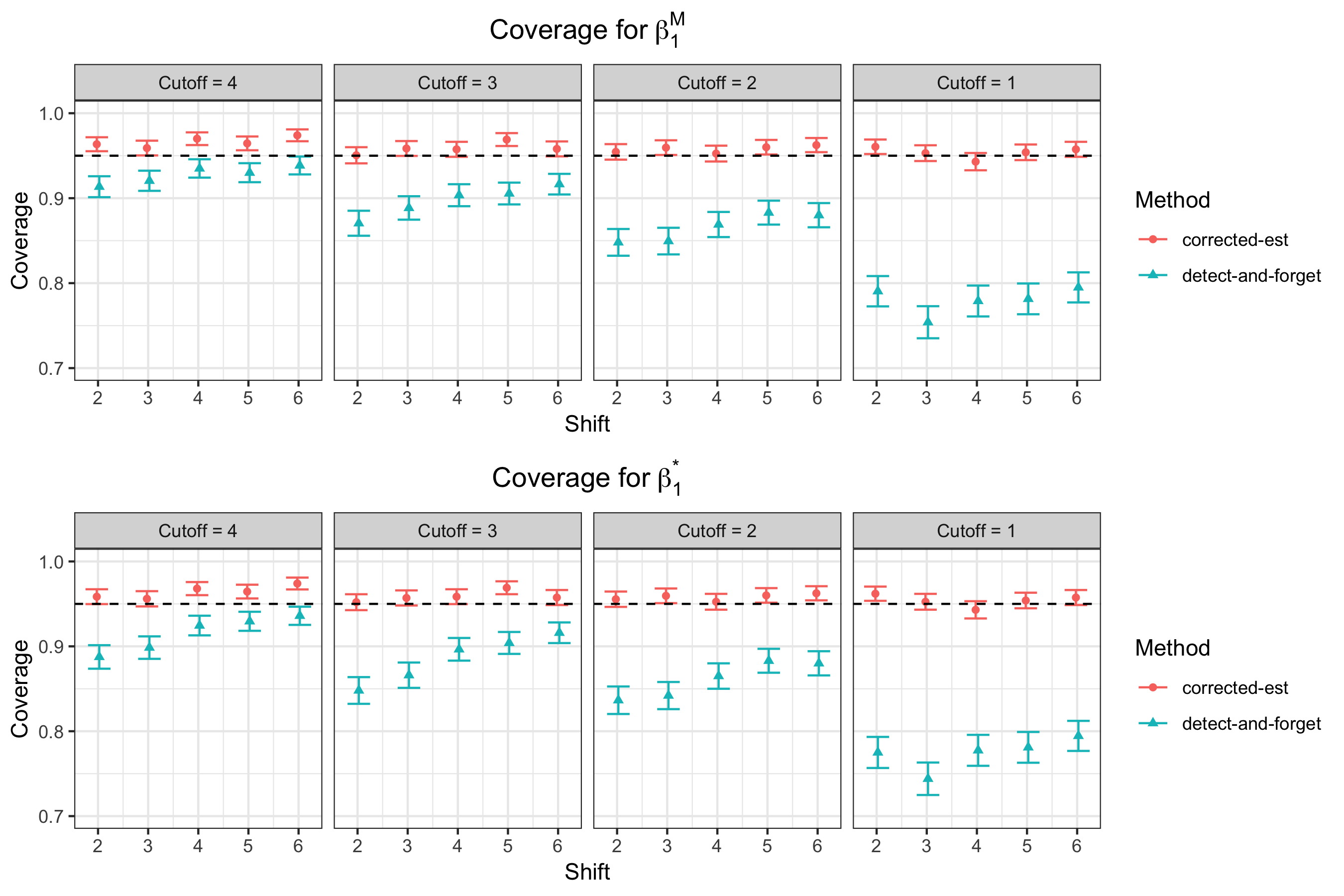}}
\caption{
Empirical coverage probability for $\bbeta^M_1$ and $\bbeta^*_1$. 
The error bar of coverage probability is obtained by $\hat q  \pm 1.96 \sqrt{\hat q (1-\hat q)/\text{nsim}}$, where $\hat q$ is the empirical coverage probability and $\text{nsim} = 2000$ is the number of realizations.
The dashed line represents $95\%$ coverage.
}
\label{fig:unknown_cov_beta1}
\end{figure}

To examine the coverage of confidence intervals for $\bbeta^M_1$, we let $\bbeta^* = (1, 2, 1, \hdots, 1)^T$ and $M^{*c} = \{1, 2, 3, 4, 5\}$.
We then fix $\bu^*_{M^{*c}} = (s, s, s, -s, -s)^T$, and we vary $s \in \{2, 3, 4, 5, 6\}$.
Outliers are then detected using Cook's distance with different cutoffs $\lambda \in \{1, 2, 3, 4\}$ as introduced in Equation \eqref{eq:cook_out_det}.
For each configuration, we do the following $2000$ times: we generate the response $\by = X \bbeta^* + \bu^* + \bvarepsilon$, where $\bvarepsilon \sim N(\b0, \sigma^2 I_n)$; we then detect outliers and form confidence intervals.
The \old~confidence intervals are set to be
$[\bnu_{\text{coef}}^T \by  \pm  \hat \sigma_\text{REFIT}  \| \bnu_{\text{coef}}  \|_2 t^{1-\alpha/2}_{|M|-p}]$,  
where $\hat \sigma_\text{REFIT}^2 =  \| \by_M - X_{M, \bigcdot} \hat \bbeta^M \|_2^2 / (|M| - p)$ (note that $\hat \sigma_\text{REFIT}$ is different from $\hat \sigma_\text{EST}$ and, as noted in \citealt{fithian2014optimal}, is generally not considered a good estimate of $\sigma$).
Figure \ref{fig:unknown_cov_beta1} shows the empirical coverage probability for $\bbeta^M_1$ and $\bbeta^*_1$. 
As our theories predict, \newest~intervals give $95\%$ coverage of $\bbeta^M_1$, while \old~intervals are off. 
Although without theoretical guarantees, \newest~intervals still achieve the desired coverage for $\bbeta^*_1$.

\begin{figure}[ht]
\centering
\includegraphics[width = \textwidth]{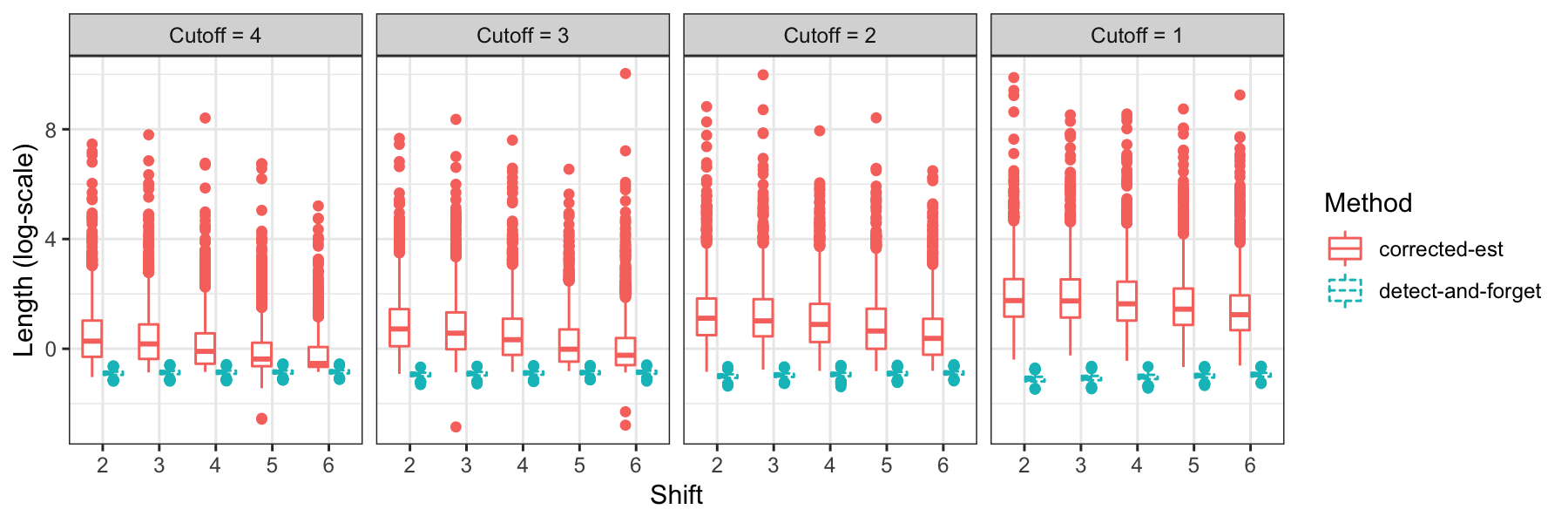}
\caption{
  Length of confidence intervals for $\bbeta^M_1$.
}
\label{fig:unknown_length_ci_beta1}
\end{figure}

Figure \ref{fig:unknown_length_ci_beta1} shows the length of both kinds of intervals. 
We see that the achievement of desired coverage comes with a price: the length of \newest~intervals is in general wider than \old~intervals.

\begin{figure}[ht]
\centering
\makebox{\includegraphics[width = \textwidth]{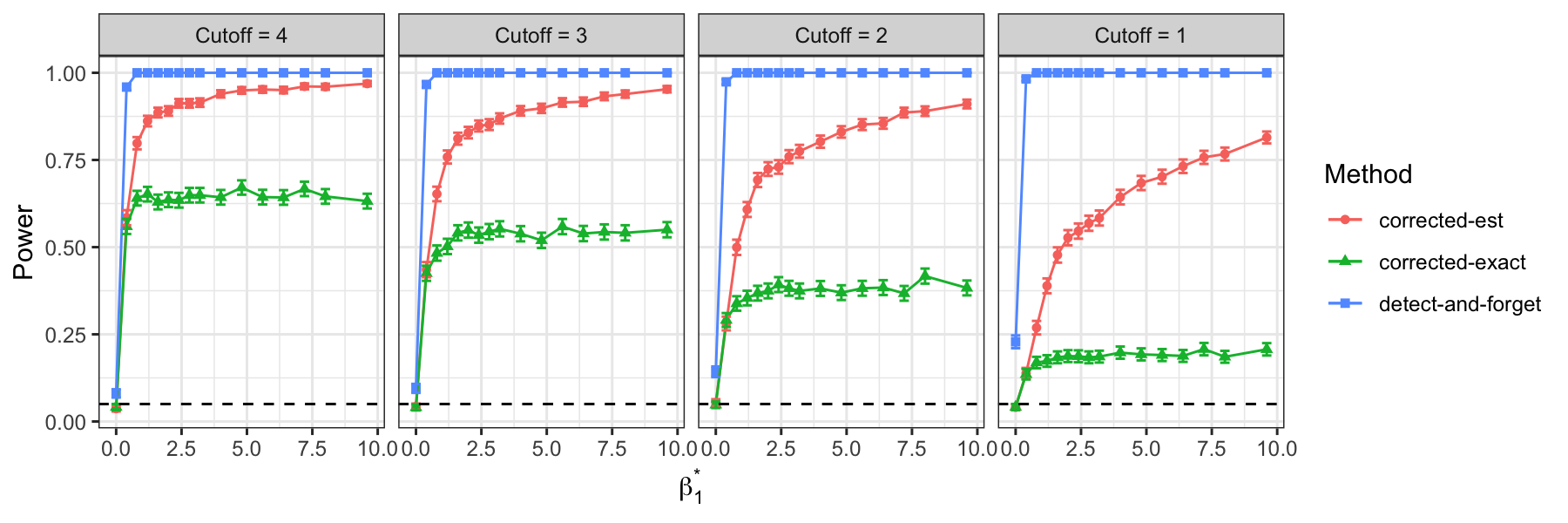}}
\caption{
  The empirical power of testing $H_0(*, 1): \bbeta^*_1 = 0$ against $H_1(*, 1): \bbeta^*_1 \neq 0$. The error bar is obtained by $\hat q  \pm 1.96 \sqrt{\hat q (1-\hat q)/\text{nsim}}$, where $\hat q$ is the empirical power and $\text{nsim} = 2000$ is the number of realizations.
}
\label{fig:unknown_power_beta1}
\end{figure}

We next examine the power of testing $H_0(M, 1): \bbeta^M_1 = 0$ against $H_1(M, 1): \bbeta^M_1 \neq 0$.
We let $\bbeta^*_k = 1$ for $k = 0, 2, 3, \hdots, 10$, and we vary $\bbeta^*_1$ smoothly.
We let $s = 4$ and the rest of the setup is the same as the previous simulation.
We run $2000$ iterations. In each iteration, we generate the response, detect outliers, and extract $p$-values.
The \old~$p$-value is set to be
$2F_t^{|M| - p}(-|\frac{\bnu_{\text{coef}}^T \by}{\hat \sigma_\text{REFIT} \| \bnu_{\text{coef}} \|_2}|)$,
where $F_t^{df}$ is the CDF of a $t_{df}$ distribution.
For power considerations, \newest~and \newexact~$p$-values and are defined as $2\min(1-\text{pval}, \text{pval})$, where $\text{pval}$ is the $p$-value calculated by directly applying Corollary \ref{cor:pw_inf_lincomb_mu_pvalue} or Theorem \ref{thm:sel_F_test}.
By construction, we are actually examining the power of testing $H_0(*, 1): \bbeta^*_1 = 0$ against $H_1(*, 1): \bbeta^*_1 \neq 0$.
Figure \ref{fig:unknown_power_beta1} shows the results: the two selective methods control the type I error down to $0.05$ even though this correspond to $H_0(*, 1)$ (recall that our theory ensures control under $H_0(M, 1)$), while \old does not.
Both \newest~and \newexact~suffer from a loss of power, although comparing to the power of \old~is not meaningful since it does not control Type I error.
The power for \newest~seems acceptable, while \newexact~has quite a substantial loss in power, which may be the consequence of conditioning on too much information.

\begin{figure}[ht]
\centering
\makebox{\includegraphics[width = \textwidth]{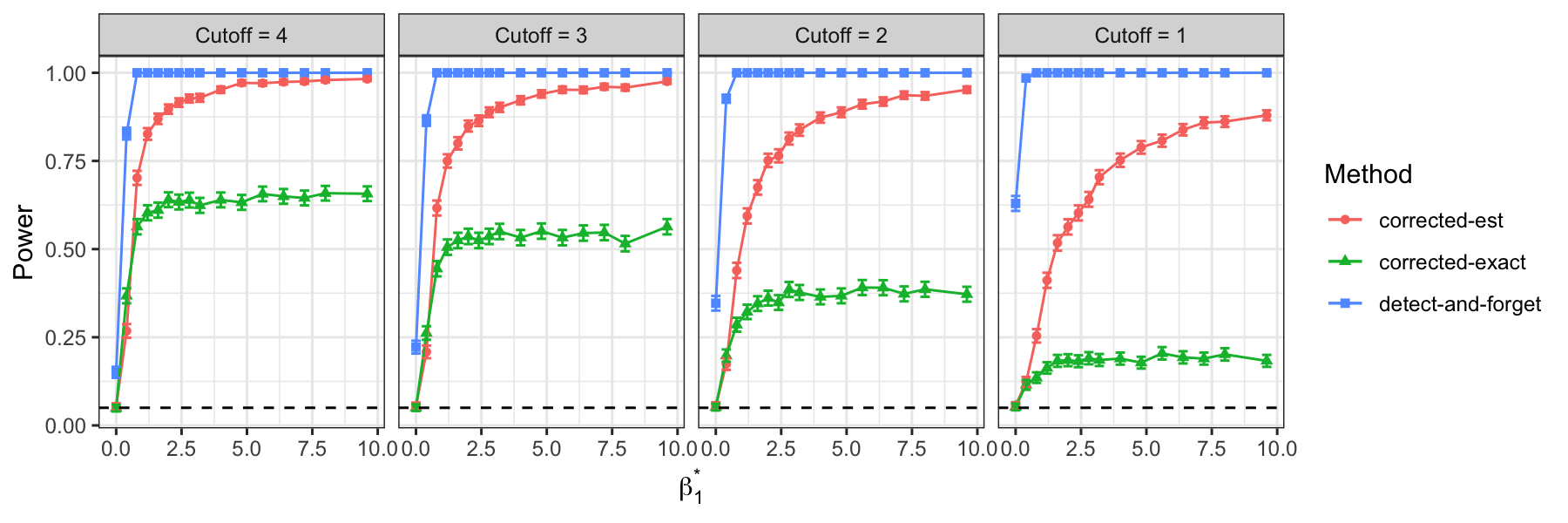}}
\caption{
  The empirical power of testing $H_0(*, g): \bbeta^*_g = \b0$ against $H_1(*, g): \bbeta^*_g \neq \b0$,The error bar is obtained by $\hat q  \pm 1.96 \sqrt{\hat q (1-\hat q)/\text{nsim}}$, where $\hat q$ is the empirical power and $\text{nsim} = 2000$ is the number of realizations.
  }
\label{fig:unknown_power_grp}
\end{figure}

We next examine the Type I error and power of testing the global hypothesis $H_0(M, g): \bbeta^M_g = \b0$ against $H_1(M, g): \bbeta^M_g \neq \b0$, where $g = \{1, 2, \hdots, 10\}$.
The setup is the same as the previous simulation, except that we let $\bbeta^*_0 = 1$, $\bbeta^*_k = 0$ for $k = 2, 3, \hdots, 10$ and we vary $\beta^*_1$ smoothly.
The \old~$p$-value is set to be
$
  1 - { F_F^{|g|,|M|-p}(\mathcal{F})},
$
where $\mathcal{F}$ is defined in Equation \eqref{eq:sel_F_test_def_of_F} and $F_F^{df1,df2}$ is the CDF of an $F_{df1, df2}$ distribution.
We examine the power of testing $H_0(*, g): \bbeta^*_g = \b0$ against $H_1(*, g): \bbeta^*_g \neq \b0$.
Figure \ref{fig:unknown_power_grp} shows the power as a function of $\bbeta^*_1$. 
Again we notice the failure of \old~method to control the Type I error and the loss of power of the two selective methods. 

\subsection{Data Examples}\label{subsec:R_package}
We next apply our method on three data sets. The first one is a data set from real estate economics, which has $n = 7820$ and $p = 17$ \citep{eichholtz2010doing}. And we apply both \newest~and \newexact~to this data set. The other two data sets are classical data sets from the outlier detection literature. Since the number of observations of these two data sets are relatively small, the estimation of $\sigma$ is not be as accurate as in previous simulations, and we only use \newexact, despite the fact that we may suffer from a substantial loss of power. 

{\noindent \bf Green Rating Data}

This data set consists of $p = 17$ covariates of $n = 7820$ buildings with places to rent \citep{eichholtz2010doing}. The covariates include area of the rental space of the building, the age of the building, and employment growth rating in the building's geographic region. Due to the page limit, we refer the readers to Section 7 of the online supplementary material for a detailed analysis of this data set, and we only present some highlights here. The most interesting covariate among all $17$ covariates is {\tt green\_rating}, ``an indicator for whether the building is either LEED- or Energystar-certified'', i.e., whether the building is a certified \emph{green building}. A green building is environmentally responsible and resource-efficient throughout its life-cycle \citep{kibert2016sustainable}. The question investigated in \cite{eichholtz2010doing} was the effect of {\tt green\_rating} on the rent charged to tenants in the building. 

To do so, we fit a linear model using log-rent as the response and assess the $p$-value associated with {\tt green\_rating}. After inspecting the diagnostic plots, we find that the naive fit (without removing any outliers) gives a model that highly violates the usual linear model assumptions (i.e., normality and homoscedasticity). Hence we use Cook's distance (with cutoff $\lambda = 4$) to detect outliers. This outlier detection procedure identifies $390$ potential outliers, and in view of the sample size, the removal of them should lead to a minimal loss of efficiency. After outlier removal, we consider three methods for constructing $p$-values: \old, \newest, and \newexact. While the \old~method gives an over-optimistic $p$-value of $4.75\times 10^{-6}$, \newest~gives $p = 0.039$ and \newexact~gives $p = 4.31\times 10^{-4}$. While the two \emph{corrected} methods give the same conclusion at $\alpha = 0.05$, the associated $p$-values are much larger than that given by the naive \old~method. The $p$-value given by \newexact~ is smaller than that given by \newest. This appears to be due to the lasso regression over-estimating $\sigma$ in this application.

{\noindent \bf Stack Loss Data}

Brownlee's Stack Loss Plant Data \citep{brownlee1965statistical} involves measures on an industrial plant's operation and has $21$ observations and three covariates. 
According to {\tt ?stackloss} in {\tt R} \citep{rlanguage}, ``{\tt Air.Flow} is the rate of operation of the plant, {\tt Water.Temp} is the temperature of cooling water circulated through coils in the absorption tower, and {\tt Acid.Conc} is the concentration of the acid circulating, minus 50, times 10.''  
The response, {\tt stack.loss}, ``is an inverse measure of the overall efficiency of the plant.'' 
This data set is considered by many papers in the outlier detection literature \citep{daniel1999fitting,atkinson1985plots,hoeting1996method}. 
The general consensus is that observations $1, 3, 4$ and $21$ are outliers.

We use Cook's distance to detect outliers, then fit the model and extract $p$-values, assuming $\sigma$ is unknown. 
The results are shown in Table \ref{table:stack_loss}. 
We see that as we detect outliers, the adjusted $R^2$ increases, which is an indication that the model is getting better. 
We also notice that \newexact~$p$-values are in general different from \old~ones, but there are several cases where the methods' $p$-values coincide (e.g. {\tt Water.Temp} with cutoff $4$). 
This is because the truncation set for the $F$ statistic is $[0, \infty)$ in those cases.  
This means that outlier removal does not have an effect on the conditional distribution of these test statistics.

\begin{table}
\centering
\tabcolsep=0.11cm
\footnotesize
\caption{\label{table:stack_loss}Inference for each variable in Stack Loss data after outlier detection using Cook's distance. The $p$-values in bold font are the selective ones, while the other $p$-values are refitted ones.}
\begin{tabular}{llllll}
& Full Fit & Cutoff = 4 & Cutoff = 3 & Cutoff = 2 & Cutoff = 1 \\ \hline
Outlier Detected  &  None  &   $21$    &     $1, 21$       &    $1, 3, 4,21$    &   $1,2,3,4,7,12,17,21$   \\ \hline
Adjusted $R^2$ &  $0.8983$ &   $0.9392$  &  $0.9171$  & $0.9692$ & $0.9057$  \\ \hline
{\tt Air.Flow}: Estimate   &  $0.7156$  &  $0.8891$   & $0.8458$ & $0.7977$ & $0.6666$ \\ \hline
{\tt Air.Flow}: $p$-value  & $5.8 \times 10^{-5}$ 
&  
\begin{tabular}[c]{@{}l@{}} $\bf 0.00403$  \\  $1.31 \times 10^{-6}$\end{tabular}  
& 
\begin{tabular}[c]{@{}l@{}} $\bf 0.345$ \\  $7.7 \times 10^{-6}$\end{tabular} 
&
\begin{tabular}[c]{@{}l@{}} $\bf 3.18 \times 10^{-4}$ \\  $2.48 \times 10^{-8}$\end{tabular}
&
\begin{tabular}[c]{@{}l@{}} $\bf 0.245$ \\  $1.19 \times 10^{-4}$\end{tabular}
\\ \hline
{\tt Water.Temp}: Estimate & $1.2953$  &  $0.8166$  & $0.8153$  &  $0.5773$  & $0.6357$  \\ \hline
{\tt Water.Temp}: $p$-value  &  $0.00263$  & 
\begin{tabular}[c]{@{}l@{}} $\bf 0.02309$ \\  $0.02309$\end{tabular}
& 
\begin{tabular}[c]{@{}l@{}} $\bf 0.335$ \\  $0.02431$\end{tabular}
&
\begin{tabular}[c]{@{}l@{}} $\bf 0.00694$ \\  $0.00408$\end{tabular}
&
\begin{tabular}[c]{@{}l@{}} $\bf 0.792$ \\  $0.01465$\end{tabular}
\\ \hline
{\tt Acid.Conc}: Estimate & $-0.1521$ & $-0.1071$ & $-0.0881$ & $-0.0671$ &  $-0.0411$             \\ \hline
{\tt Acid.Conc}: $p$-value & $0.34405$  
& 
\begin{tabular}[c]{@{}l@{}} $\bf 0.40234$ \\  $0.40234$\end{tabular}
&
\begin{tabular}[c]{@{}l@{}} $\bf 0.376 $ \\  $0.49585$\end{tabular}
&
\begin{tabular}[c]{@{}l@{}} $\bf 0.2961$ \\  $0.2961$\end{tabular}
& 
\begin{tabular}[c]{@{}l@{}} $\bf 0.208$ \\  $0.653$\end{tabular}
\\ \hline           
\end{tabular}
\end{table}

{\noindent \bf Scottish Hill Races Data}

This data set records the time for $35$ Scottish hill races in 1984 \citep{atkinson1986influential}. 
There are two covariates: ``{\tt dist} is the distance in miles, and {\tt climb} is the total height gained during the route in feet''. 
The response, {\tt time}, ``is the record time in hours''. 
This data set is also a classic one considered by many papers in the outlier detection literature \citep[e.g.,][]{atkinson1986influential,hadi1990stepwise,hoeting1996method}. 
The consensus is that observation 7 and 18 are obvious outliers, while observation 33 is an outlier that is masked by the other two outliers.

Again, we use Cook's distance to detect outliers, then fit the model and extract $p$-values, assuming $\sigma$ is unknown. 
The results are shown in Table \ref{table:scottish_race}. 
We can see the increase in adjusted $R^2$ as outliers are detected, and the \newexact~$p$-values differ from the \old~$p$-values in general. 
Observation $33$ is not detected until the cutoff is set to $1$, and observation $11$ is always detected as an outlier.  
\citet{atkinson1986influential} reports that observations $7, 18, 11, 33, 35$ are high-leverage points but argues that only $7, 11, 33$ are actual outliers, while the others are high-leverage points that agree with the bulk of the data.  
But we recall that our intent is not to concern ourselves with the accurate detection of outliers but rather with the proper adjustment to inference based on outlier detection and removal.

\begin{table}
\centering
\tabcolsep=0.11cm
\footnotesize
\caption{\label{table:scottish_race}Inference for each variable in Scottish Hill Races data after outlier detection using Cook's distance. The $p$-values in bold font are the selective ones, while the other $p$-values are refitted ones.}
\begin{tabular}{llllll}
& Full Fit & Cutoff = 4 & Cutoff = 3 & Cutoff = 2 & Cutoff = 1 \\ \hline
Outlier Detected &  None  &   $7, 11, 18$    &     $7, 11, 18$       &    $7, 11, 18, 31$    &   $7, 11, 18, 31, 33, 35$   \\ \hline
Adjusted $R^2$ &  $0.914$ &   $0.9721$  &  $0.9721$  & $0.9723$ & $0.9395$  \\ \hline
{\tt dist}: Estimate   &  $0.1036$  &  $0.1138$   & $0.1138$ & $0.1111$ & $0.1034$ \\ \hline
{\tt dist}: $p$-value  & $9.94 \times 10^{-12}$ 
&  
\begin{tabular}[c]{@{}l@{}} $\bf 1.76 \times 10^{-6}$  \\  $6.80 \times 10^{-15}$\end{tabular}  
& 
\begin{tabular}[c]{@{}l@{}} $\bf 1.06 \times 10^{-4}$ \\  $6.80 \times 10^{-15}$\end{tabular} 
&
\begin{tabular}[c]{@{}l@{}} $\bf 0.1219$ \\  $6.50 \times 10^{-14}$\end{tabular}
&
\begin{tabular}[c]{@{}l@{}} $\bf 6.99 \times 10^{-9}$ \\  $2.13 \times 10^{-13}$\end{tabular}
\\ \hline
{\tt climb}: Estimate & $1.84 \times 10^{-4}$  &  $1.28 \times 10^{-4}$  & $1.28 \times 10^{-4}$  &  $1.42 \times 10^{-4}$  & $1.17 \times 10^{-4}$  \\ \hline
{\tt climb}: $p$-value  &  $6.49\times 10^{-6}$  & 
\begin{tabular}[c]{@{}l@{}} $\bf 0.05918$ \\  $9.15\times 10^{-6}$\end{tabular}
& 
\begin{tabular}[c]{@{}l@{}} $\bf 0.02465$ \\  $9.15\times 10^{-6}$\end{tabular}
&
\begin{tabular}[c]{@{}l@{}} $\bf 0.06060$ \\  $1.16\times 10^{-5}$\end{tabular}
&
\begin{tabular}[c]{@{}l@{}} $\bf 7.02\times 10^{-4}$ \\  $1.53 \times 10^{-5}$\end{tabular}
\\ \hline         
\end{tabular}
\end{table}

\section{Discussion}\label{sec:discussion}

In this paper, we have introduced an inferential framework for properly accounting for the removal of outliers from a data set.  
The commonplace approach, \old, makes the incorrect assumption that outlier removal does not affect the distribution of the data.  
Our work is based on recent developments in the selective inference literature, which carries out inference that properly accounts for variable selection \citep{lee2016exact,loftus2015selective}.  
A key idea in that work is to characterize the event that a certain set of variables is selected in terms of a simple to describe set of constraints on the response vector $\by$.  
Doing so makes it tractable to derive the conditional distribution of the estimator given this selection event.  
Our work likewise relies on the fact that the most commonly used outlier detection procedures can be expressed in a relatively simple form, namely a quadratic constraint on the response vector. Our results can be in principle extended to ``convex detection procedures'', where the event $\{\widehat M = M\}$ is characterized as a convex constraint on the response, using the results from \cite{harris2016selective}.

Our target of inference is $\bbeta^M$, where $M$ is the selected set
of non-outliers. By focusing on  $\bbeta^M$, we are able to decouple
the challenge of identifying outliers from the focus of our work,
which is accounting for the search and removal of potential outliers.
When $M$ excludes all true outliers, $\bbeta^M$ coincides with $\bbeta^*$.  
When the true outliers are easily detected, then (via Proposition \ref{prop:unconditional_coverage}), our methodology translates to inference on $\bbeta^*$.  
However, when there are true outliers that are undetected, our
statements about $\bbeta^M$ may not translate well to statements on
$\beta^*$.  In some cases, an outlier may not be too severe and
therefore go undetected; in such a case, $\bbeta^M$ would not be too far from $\beta^*$, in which case our inferential statements may be translated, approximately, to statements about $\bbeta^*$.  
An interesting future direction would be to characterize the regimes (in terms of size of outlier) in which (i) all true outliers are easily detected and thus we can make inferential statements about $\bbeta^*$ and (ii) not all outliers are easily detected but $\bbeta^M\approx \bbeta^*$ so that approximate statements about $\bbeta^*$ can be made.  
And of course, a central question would then be whether there is a gap between regimes (i) and (ii).

The inferential framework introduced in this paper suffers from a loss of power, especially in the case of unknown $\sigma$. A possible remedy is to introduce some randomization into the outlier detection procedure. For example, one can adopt the strategy of \cite{tian2018selective}, namely adding a properly scaled Gaussian noise to the response, so that the selective tests can have a better power at the cost of a less accurate outlier detection procedure. Investigating possible strategies to increase the power remains for future work.

In this paper, we have provided frequentist inference in the linear model after outlier removal. However, with the characterization of the detection procedure at hand, our method can be extended to a Bayesian setup, namely constructing the appropriate detection-adjusted posterior on the regression coefficients, by adapting the results from \cite{yekutieli2012adjusted,panigrahi2016bayesian}.

Another future direction would be to consider proper inference after outlier removal in the high-dimensional setting.
Our method explicitly assumes a low-dimensional setting through the assumption that $X_{\hM, \bigcdot}$ has linearly independent columns.
A direct generalization of the outlier detection method \eqref{eq:ipod} is to instead solve
\[
  (\hat \bbeta, \hat \bu) = \underset{\bbeta \in \mathbb{R}^p, \bu \in \mathbb{R}^n}{\arg \min} \frac{1}{2n} \| \by - X\beta - \bu\|_2^2 + \lambda_1 \|\bbeta\|_1 + \lambda_2 \| \bu\|_1.
\]
Applying \citet[Theorem 4.3]{lee2016exact}, one could perform inference corrected simultaneously for both variable selection and outlier removal. 
Another approach would be to use a high-dimensional extension of Cook's distance proposed by \citet{zhao2013high} (it too can be shown to be a quadratic outlier detection procedure). 
One could then do variable selection with the remaining data, for example using the lasso. In this case our methodology would still, in principle, apply.
Characterizing the exact conditional distributions from more general
procedures, such as after MM-estimation, remains a non-trivial problem.

\section*{Acknowledgements}
The authors gratefully acknowledge support from an NSF CAREER grant, DMS-1653017, and thank an associate editor for pointing us to the green rating data set.

\bigskip
\begin{center}
{\large\bf SUPPLEMENTARY MATERIAL}
\end{center}

\begin{description}

\item[Supplementary material for this manuscript:] For brevity, we collect proofs of most theoretical results, some additional simulation results and the implementation details in the online supplementary material. (available at \url{https://www.dropbox.com/s/o9xxkap0q68knbc/supplements.pdf?dl=0}) 

\item[{\tt R} package {\tt outference}:] {\tt R} package containing code to perform the inferential methods described in this paper. (available at \url{https://github.com/shuxiaoc/outference})

\item[{\tt R} scripts] {\tt R} scripts to reproduce all figures and simulation results in this paper. (.zip file)

\end{description}

{
\bibliographystyle{chicago}
\bibliography{ref}
}

\end{document}